\documentclass[twocolumn,showpacs,prb,preprintnumbers]{revtex4}
\usepackage{graphicx}
\usepackage{dcolumn}
\usepackage{bm}

\preprint{(PhysicaC) in press}

\begin{document}
\title{Effects of unreacted Mg impurities on the transport properties of MgB$_{2}$}
\author{C. U. Jung}
 \altaffiliation[New address: ]{Tokura Spin Superstructure Project,
 ERATO, JST, AIST, Tsukuba Central 4,
 Tsukuba, Ibaraki 305-8562, Japan}
 \email[E-mail: ]{cu-jung@aist.go.jp}
 \homepage[Url: ]{http://unit.aist.go.jp/cerc/}
\author{Heon-Jung Kim}
\author{Min-Seok Park}
\author{Mun-Seog Kim}
\author{J. Y. Kim}
\author{Zhonglian Du}
\author{Sung-Ik Lee}
\affiliation{National Creative Research Initiative Center for
Superconductivity and Department of Physics, Pohang University of
Science and Technology, Pohang 790-784, Republic of Korea }
\author{K. H. Kim}
\author{J. B. Betts}
\author{M. Jaime}
\author{A. H. Lacerda}
\author{G. S. Boebinger}
\affiliation{NHMFL, Los Alamos National Laboratory, MS E536, Los
Alamos, NM 87545 USA}

\date{\today }
\begin{abstract}
We synthesized polycrystalline MgB$_{2}$ from a stoichiometric
mixture of Mg and the $^{11}$B isotope under different conditions.
All the samples showed bulk superconductivity with $T_{c}=38\sim
39$ K. The samples containing the least amount of unreacted Mg
showed the highest $T_{c}$ and the sharpest
transition width ($\Delta T_{c}$). A residual resistivity ratio (RRR) of $%
\sim 5.8$, and a magnetoresistance (MR),\ at 40 K, of 12\% were
obtained for these samples. Moreover, there was no upturn of
resistivity in a low temperature region at 10 Tesla. The samples
containing appreciable amounts of unreacted Mg showed quite
different behaviors; the values of $\Delta T_{c} $, RRR,\ and MR
were much larger. An upturn appeared in resistivity of the samples
below about 50 K at 10 T and is thought to be due to the unreacted
Mg.
\end{abstract}

 \maketitle

The recent discovery\cite{find} of the binary metallic MgB$_{2}$
superconductor has attracted great scientific\cite
{Canfield,Budko,KangHall,BudkoPRB,Finnemore,ChYeh,Vasquez} and
industrial \cite{Canfield,Kangfilm} interest. Even though basic
issues, such as the isotope effect\cite{Budko} and the
determination\cite{KangHall} of the carrier type, were addressed
immediately, there still exist conflicting results. For example,
several groups\cite
{Canfield,BudkoPRB,Finnemore,JungC353,Chu,Zhu,Frederick1,Frederick2,Fuchs}
have measured the transport properties of their polycrystalline
MgB$_{2}$ and have reported different values for the residual
resistivity ratio (RRR), magnetoresistance (MR), residual
resistivity. The widely accepted opinion was that good bulk
samples of MgB$_{2}$ (called `{\it highRRR-MgB}$_{2}$') should
have higher values of the RRR ($20\sim 25$) with low residual
resistivity ($\rho (40$ K$)=0.38-1$ $\mu \Omega $cm) and higher MR
with a resistivity upturn at low temperature under high magnetic
field.\cite {Canfield,BudkoPRB,Finnemore} Insulating impurities
and/or local strains were thought to be the causes of any
behaviors different from the above.\cite
{JungC353,Chu,Zhu,Klie,0598} However, recent reports on all single
crystals \cite{SingleKim,Lee,Xu,Karpinski} and many polycrystals
\cite {Frederick1,Frederick2,Fuchs,SWCheong} have shown a RRR
value of about 5.
Moreover, the residual resistivity of single crystals ($\rho _{ab}(40$ K$%
)=1-2$ $\mu \Omega $cm) was higher than that of `{\it
highRRR-MgB}$_{2}$'. \cite{SingleKim,Lee,Xu,Karpinski} Thus, it
has become an urgent matter to clarify the origin of the different
reports on polycrystalline MgB$_{2}$.

Here, we report the transport properties of MgB$_{2}$ prepared
under different conditions. For samples containing the least
amount of unreacted Mg, we found $T_{c}\simeq 39.2$ K, $\Delta
T_{c}=0.3-0.5$ K, RRR $\sim 5.8$, and MR$_{5T}\sim 3$\%. The last
three values are very similar to those reported for single
crystals. For the samples containing a relatively large amount of
unreacted Mg, the values for $\Delta T_{c}$, RRR, and MR, were
considerably larger. These higher values of RRR and MR are simlar
to those in {\it highRRR-MgB}$_{2}$. $T_{c}$\ for these samples
was about 38 K.

\begin{figure}[tbp]
\includegraphics[width=7.0cm]{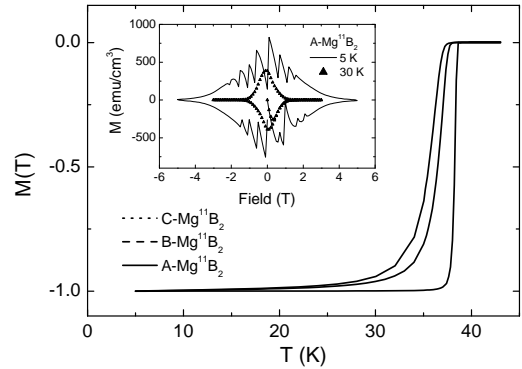}
\caption{Normalized magneization of MgB$_{2}$ measured at 20 Oe.
The solid line, the dashed, and the dotted lines are for
A-MgB$_{2}$, B-MgB$_{2}$, and C-MgB$_{2}$, respectively. $T_{c}$
was $38\sim 39$ K and $\Delta T_{c}$ was 0.5, 3.5, and 4.5 K,
respectively. The inset shows the magnetic hysteresis curve,
$M$($H$), of a piece of A-MgB$_{2}$ (m $\sim 78$ mg) measured at 5
K and 30 K.} \label{MT}
\end{figure}

A Ta capsule containing a stoichiometric mixture of Mg chunks and the $^{11}$%
B isotope was heated under 3 GPa in a 12-mm cubic multi-anvil-type
press.
\cite{JungC353,JungAPL,Jung6441,JungMCNS} The maximum heating temperature ($%
850\sim 1000^{\circ }$C) and the preparation of the precursors
were varied to obtain different samples. One batch, A-MgB$_{2}$,
was heated at $950\sim 1000^{\circ }$C after enough grinding of
the precursors. For B-MgB$_{2}$ and
C-MgB$_{2}$, the heating temperatures were about 900$^{\circ }$C and $%
850^{\circ }$C\ respectively, and the precursors were ground less.
Details of the high pressure synthesis have been reported
previously.\cite {JungC353,JungAPL} The pellet density was
$2.48\sim 2.6$ g/cm$^{3}$ which is quite close to the
crystallographic density, a common feature in high-pressure
synthesis.\cite{JungC353,TakanoAPL} As a result, grain
connectivity was prevalent over the entire sample ($\sim 1$ mm).

A dc SQUID magnetometer (Quantum Design, MPMS{\it XL}), a
field-emission scanning electron microscope (SEM), an optical
microscope, and an X-ray diffractometer were used for this
investigation. The resistivity, $\rho (T,H) $, was measured by
using a standard 4-probe technique on a bar shaped specimen ($\sim
0.5\times 1\times 4$ mm$^{3}$).

Figure \ref{MT} shows the normalized magnetization curves $M(T)$
measured at
20 Oe in the zero-field cooled mode for A-MgB$_{2}$, B-MgB$_{2}$, and C-MgB$%
_{2}$. All samples showed $M/H\gtrsim 150\%$ of $-1/4\pi $, as
previously
observed.\cite{Budko} The $T_{c}$ of A-MgB$_{2}$ was about 39 K, and its $%
\Delta T_{c}$ 0.5 K. The values of $T_{c}$ for B-MgB$_{2}$ and
C-MgB$_{2}$ were slightly lower, but the transition width $\Delta
T_{c}$ became much larger, i.e., 3.5 and 4.5 K,
respectively.\cite{discriminate} The inset in Fig. \ref{MT} shows
the magnetic hysteresis curves $M$($H$) for A-MgB$_{2}$ at 5 K and
30 K; a very clear flux jumping behavior was observed at 5 K. This
kind of flux jump has also been reported for commercial MgB$_{2}$
(HIP-Alfa) after sintering under 3 GPa.\cite{MSKimPRB} Details of
this result will be reported elsewhere.\cite{MSKim}

\begin{figure}[tbp]
\includegraphics[width=6.5cm]{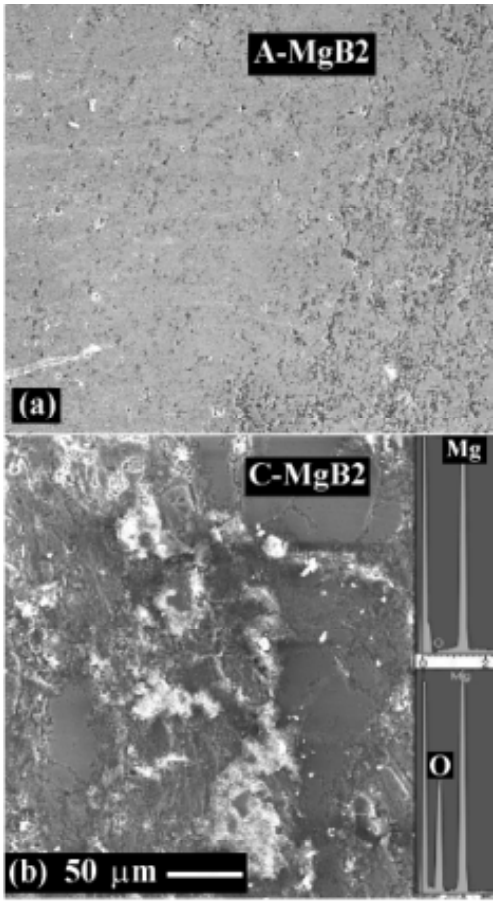}
\caption{SEM images of A-MgB$_{2}$ and C-MgB$_{2}$ at the same
magnification with scale bar 50 $\protect\mu $m in length: (a)
well connected grains in
A-MgB$_{2}$, (b) Unreacted Mg in C-MgB$_{2}$, as large as several tens of $%
\protect\mu $m, are shown as dark islands. For example, see an
island under caption of `C-MgB2'. } \label{SEM}
\end{figure}

To search for the origin of the above difference in $M(T)$, we
investigated images of cleaved or polished surfaces by using a SEM
and an optical microscope. The SEM image for A-MgB$_{2}$ in Fig.
\ref{SEM}(a) shows well-connected grains, with no clear grain
boundaries, over a very wide region. However, unreacted Mg was
easily identified for C-MgB$_{2}$, as
shown in Fig. \ref{SEM}(b). The dark islands, as large as several tens of $%
\mu $m, in Fig. \ref{SEM}(b) were identified to be unreacted Mg by
a clear Mg peak at 1.25 keV in the energy dispersive spectrum
(EDS) in the inset of Fig. \ref{SEM}(b). For example, see an
island under caption of `C-MgB2'.
Another evidence is that strong Mg peak in XRD pattern such as one near $%
2\theta \sim 37^{\circ }$\ reported previously\cite{Zhu} was also
observed for C-MgB$_{2}$. Unreacted Mg grains were also found in
B-MgB$_{2}$, but were nearly absent in A-MgB$_{2}$. The unreacted
Mg grains in C-MgB$_{2}$ were bigger and had denser population
than those in B-MgB$_{2}$. The unreacted Mg showed a
characteristic white tint when viewed through the optical
microscope. All these shows that the basic superconducting
parameters $T_{c}$\ and $\Delta T_{c}$\ were degraded as the
amount of unreacted Mg is increased. For C-MgB$_{2}$, insulating
MgO impurity was also identified by a clear O peak at 0.52 keV in
EDS and was shown as white dots in Fig. \ref{SEM}(b).

\begin{figure}[tbp]
\includegraphics[width=7cm]{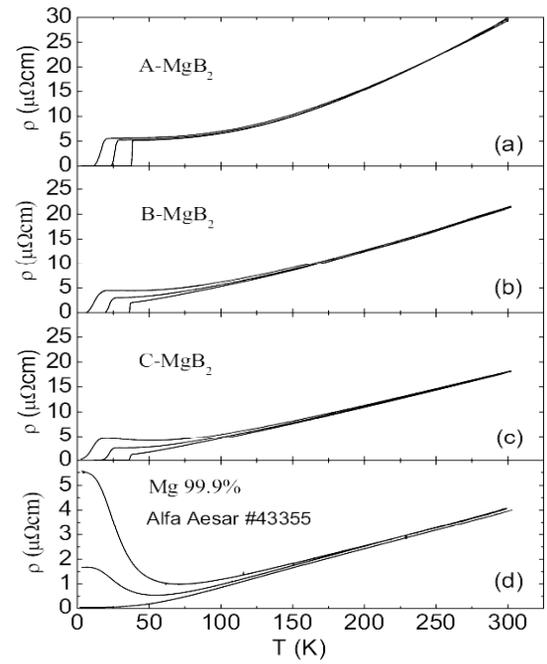}
\caption{Resistivity, $\protect\rho $($T$), for various kinds of
MgB$_{2}$ at 0, 5, and 10 Tesla. (a) The $T_{c}$ was about 39 K,
and the $\Delta T_{c}$ was 0.3 K, the RRR = 5.8 and the values of
the MR at 40 K were 3\% and 12\% at 5 and 10 Tesla for
A-MgB$_{2}$. (b) For B-MgB$_{2}$, RRR = 9.9, and the values of the
MR at 40 K were 49\% and 111\%. (c) For C-MgB$_{2}$, RRR = 13.7
and the values of the MR at 40 K were 100\% and 224\%. (d)
Resistivity for Mg (Alfa Aesar \#43355). Note that the resistivity
upturn below about 50 K became more pronounced at higher fields
for C-MgB$_{2}$\ and the Mg.} \label{RTH}
\end{figure}

To further investigate the effect of unreacted Mg, we measured
resistivity under a magnetic field. Figure \ref{RTH} shows $\rho
(T)$ at 0, 5, and 10 T. A-MgB$_{2}$ shows the highest $T_{c}$,
$\sim 39$ K, and the smallest $\Delta
T_{c}$, $\sim 0.3$ K. The values of the zero-field resistivity for A-MgB$%
_{2} $ at 40 K and 300 K were 5.13 and 30.0 $\mu \Omega $cm,
respectively. A RRR value of $\sim 5.8$ and a MR of $3\%(12\%)$ at
5(10) T were also observed. The RRR for all stoichiometric samples
prepared under similar
conditions had values from $\sim 4.5$ to $\sim 6$. The values of RRR and MR$%
_{5T}$ were similar to those recently reported for single crystal
\cite {SingleKim,path} and
polycrystal\cite{Frederick,Fuchs,SWCheong}.

While $\rho (300$ K) was not much different, the $\rho (40$ K$)$ of $%
2.11(1.29)$ $\mu \Omega $cm for B-MgB$_{2}$(C-MgB$_{2}$) was much
smaller than that of A-MgB$_{2}$, giving a larger value of RRR =
9.9(13.7). Also, the MR$_{5T}$ increased drastically up to 49\%
and 100\%, as shown in Fig. \ref{RTH}(b) and Fig. \ref{RTH} (c).
At a higher field of 10\ T, MR$_{10T}$ reached values as large as
111\%(224\%) for B-MgB$_{2}$(C-MgB$_{2}$). Another interesting
observation is the peculiar resistivity upturn below 50 K for
C-MgB$_{2}$\ at 10 T. A similar upturn had been reported
previously for {\it highRRR-MgB}$_{2}$.\cite{BudkoPRB}

Now let's turn our attentiuon to the resistivity behavior of Mg
itself. Figure \ref{RTH}(d) shows the resistivity of a pressed bar
of a commercial Mg block (Alfa Aesar \#43355) measured at 0, 5,
and 10 T. The observed temperature dependence of $\rho (T$, $H=0$)
with very high RRR value is consistent with that reported in the
literature\cite{Mgdata} for pure Mg except for the slightly higher
value of 0.12 $\mu \Omega $cm at 40 K. The resistivity of Mg under
an applied magnetic field shows several interesting features,
especially at lower temperatures, such as high MR values and a
large resistivity upturn below 50 K. The MR at 5(10) T is more
than 400\%(1300\%) at 40 K, and the resistivity upturn became more
pronounced at higher fields. Because $\rho (T)$ of Mg is much
smaller than $\rho (T)$ of
stoichiometric A-MgB$_{2}$ in the region of $T>T_{c}$ and $H\leq 10$ T, $%
\rho (T,H)$ of MgB$_{2}$ containing unreacted Mg\ will have some
of the character of Mg. The unreacted Mg in MgB$_{2}$ will
decrease the resistivity at 40 K more than it will at 300 K, as
observed in Figs. \ref{RTH}(b) and \ref{RTH}(c), so the resulting
increase in the RRR\ should be accompanied by an increase in the
MR. Moreover, the Mg in C-MgB$_{2}$ also explains the resistivity
upturn below 50 K at 10 T.

Incomplete reaction of polycrystalline MgB$_{2}$ due to either
unoptimized growth conditions or a Mg-rich precursor may result in
unreacted Mg.\cite {surfaceMg} Unlike the insulting oxide
impurities in high-$T_{c}$ cuprates, unreacted conducting Mg may
not be so detrimental in bulk transport applications of MgB$_{2}$,
especially for the case of quenching of the superconductivity.
However, if a correct understanding of the intrinsic properties of
MgB$_{2}$ such as resistivity, which can be screened by the
unreacted conducting Mg impurities, is to be obtained, then phase-pure MgB$%
_{2}$ is essential.

In conclusion, we synthesized MgB$_{2}$ and investigated
systematically the effect of unreacted Mg in the MgB$_{2}$ on its
transport properties. For the stoichiometric MgB$_{2}$, RRR was
$\sim 5.8$, and MR$_{5T}$ was $\sim 3$\%. We found that the larger
transition width and the higher values of the RRR and the MR were
due to the unreacted Mg in the MgB$_{2}$. The appearance of
peculiar resistivity upturn at a lower temperatures at higher
field of 10 T was also explained as being due to the unreacted Mg
in the MgB$_{2}$ materials. This conclusion is also supported by
the results for single crystals of MgB$_{2}$.

\acknowledgments This work is supported by the Ministry of Science
and Technology of Korea through the Creative Research Initiative
Program. We appreciate valuable discussions with S.-W. Cheong and
Y. S. Song.


\begin{references}
\bibitem{discriminate}  The $T_{c}$ of B-MgB$_{2}$ and C-MgB$_{2}$ seem to
be a little bit lower than that of A-MgB$_{2}$, which seems to
invoke some intragranular origin. As suggested in a previous
report(Ref. 11), the unreacted Mg may cause local
off-stoichiometry, resulting in a micro-strain.
\bibitem{path}  Even though the in-plane resistivity of a single crystal has
not been determined accurately($\rho (40$ K$)=1-2$ $\mu \Omega $cm
in Ref. 19 and Ref. 21.), the comparison of the RRR with that of
our polycrystalline sample($\rho (40$ K$)\sim 5$ $\mu \Omega $cm)
is reasonable if the in-plane resistivity is more conductive than
the c-axis resistivity of MgB$_{2}$, which has been generally
assumed to be the case due to its layered structure. If the
driving current flows main along the ab-plane of each grains, the
effective area will be reduced and the effective length will be
increased. All these can account for the factor of difference of
about 3. However, more rigorous study of the c-axis resistivity,
as well as the in-plane resistivity, seems to be necessary for a
clear comparison.
\end{references}
\end{document}